\documentclass[aps,prb,reprint,groupedaddress,floatfix,amsmath,amssymb]{revtex4-2}
\usepackage{graphicx}
\usepackage{dcolumn}
\usepackage{hyperref}
\hypersetup{colorlinks=true,linkcolor=blue,citecolor=blue,urlcolor=blue}

\newcommand{\ket}[1]{|#1\rangle}
\newcommand{\bra}[1]{\langle #1|}
\newcommand{\bracket}[2]{\langle #1|#2\rangle}
\newcommand{\Braket}[3]{\langle #1|#2|#3\rangle}
\newcommand{\one}{\mathbb{1}}
\newcommand{\hc}{\text{h.c.}}
\newcommand{\be}{\begin{equation}}
\newcommand{\ee}{\end{equation}}
\newcommand{\bea}{\begin{eqnarray}}
\newcommand{\eea}{\end{eqnarray}}
\newcommand{\dg}{\dagger}
\newcommand{\Dc}{\Delta_c}
\newcommand{\dt}{\delta t}\newcommand{\Dcm}{\Delta_{c,-}}\newcommand{\Dcp}{\Delta_{c,+}}
\newcommand{\Fo}{F_1}
\newcommand{\Hz}{\text{\,Hz}}

\graphicspath{{figures/}}

\begin{document}

\title{Competition and cancellation in a tilted Rice--Mele chain:
exact Wannier--Stark offset and spectral readout}

\author{Xiang Gao$^{1}$} \email[E-mail: ]{gaoxiang@zknu.edu.cn}
\author{Xing Deng$^{2}$}
\author{Shoulong Chen$^{3}$}
\author{Ying Zhang$^{1}$}
\author{Chunhua Zhang$^{1}$}
\author{Wei Xiao$^{1}$}
\author{Xueting Fang$^{4}$} \email[E-mail: ]{2024510219@hust.edu.cn}
\affiliation{$^1$School of Physics and Telecommunications Engineering, Zhoukou Normal University, Zhoukou 466001, China \\
$^2$Shiyan Key Laboratory of Quantum Information and Precision Optics, and School of Mathematics, Physics and Optoelectronic Engineering, Hubei University of Automotive Technology, Shiyan 442002, China\\
$^3$Department of Physics and Center for Computational Sciences, Sichuan Normal University, Chengdu 610066, China\\
$^4$MOE Key Laboratory of Fundamental Physical Quantities Measurement and Hubei Key Laboratory of Gravitation and Quantum Physics,
PGMF and School of Physics, Huazhong University of Science and Technology, Wuhan 430074, China\\
}


\begin{abstract}
In a tilted Rice--Mele chain, the Wannier--Stark (WS) ladder becomes staggered, \(E_n=n\Fo+(-1)^{n+1}A\). The offset \(A\) reflects the competition between the staggered onsite energy \(\Delta\) and the dimerized hopping \(\dt\), and its cancellation restores spectral equidistance.
We determine \(A\) through a systematic strong-field series in the WS basis and a nonperturbative two-band Floquet expression. An exact matrix element reduces the strong-field expansion to Bessel algebra; at the representative working point, truncation at seventh order locates the cancellation point with a theory-internal truncation error of \(0.010\)~Hz, against \(27.3\)~Hz at first order.
The Floquet propagator gives the closed form \(A=(\Fo/\pi)\arcsin(-i\,\mathrm{Tr}\,U/2)\), exact within the two-band model, implying the kinematic bound \(|A|\le\Fo/2\) and identifying cancellation with maximum spectral uniformity.
The same analytic structure yields a family of real zeros and complex exceptional points that set the convergence radius of the strong-field series without touching the real axis. On odd-length chains, the hidden-symmetry defect produces a pairing plateau of height \(2|A|\) that persists numerically beyond first order in the modulations, allowing the offset to be read from the static spectrum without Floquet propagation or time evolution.
In the bulk, cancellation appears as the merging of two sidebands into a single line at \(\Fo\), a null measurement that requires no calibrated sideband amplitudes, although the tilt and line frequency still need calibration.
These signatures provide complementary spectral tests of the staggered WS ladder and a calibration-aware route to precision force sensing.
\end{abstract}

\maketitle

\section{Introduction}

A particle in a periodic potential, subjected to a static force, undergoes Bloch oscillations (BOs), and its spectrum is quantized into an equidistant Wannier--Stark (WS) ladder $E_n=n\Fo+\mathrm{const}$ \cite{Bloch28,Zener34,Houston40,Wannier60}, observed as resonances in semiconductor superlattices~\cite{Gluck02} and as WS ladders in ultracold atoms~\cite{BenDahan96,Wilkinson96}. 
For an ideal single-band tight-binding system, this equidistance stems from the covariance of the Hamiltonian: under translation by one lattice period the Hamiltonian shifts by a constant $\Fo$; this fixes the spacing and the Bloch frequency, and it underlies the use of BOs in force sensing and interferometry~\cite{Morsch06,Gluck02}. 
In a bulk lattice, a period-2 modulation of the potential doubles the lattice constant, $a\to2a$, and halves the Brillouin zone. 
Translation by $2a$ then shifts the Hamiltonian by $2\Fo$, guaranteeing two interleaved subladders, each of spacing $2\Fo$, but leaving the relative offset of the two subladders undetermined.

A paradigmatic model for addressing this offset is the Rice--Mele model~\cite{Rice82,Thouless83,DeBeule24} with dimerized hopping $t+(-1)^n\dt$, staggered onsite energies $\pm\Delta/2$, and a static tilt $\Fo$. 
It is realized in double-well optical superlattices~\cite{SebbyStrabley06,Folling07}, where $\dt$ and $\Delta$ can be tuned independently~\cite{Atala13,Nakajima16,Lohse16,Citro23}.
In the bulk, its spectrum takes the staggered-ladder form 
\begin{equation}
    E_n=n\Fo+(-1)^{n+1}A ,
    \label{eq:ladder}
\end{equation}
that is, two interleaved subladders offset by $\pm A$, with spacings alternating between $\Fo-2|A|$ and $\Fo+2|A|$. 
Reversing the sublattice convention flips the sign of $A$ but leaves $|A|$ and all level spacings unchanged. 
The offset is thus both spectroscopically measurable and tunable through the two modulations. 

Away from cancellation, $A$ appears as a beat in BOs and as the splitting of spectroscopic sidebands~\cite{Breid06,Dreisow09,Kling10}. 
Near cancellation, the competing onsite and bond modulations drive $A$ to zero, the sidebands at $\Fo\pm2|A|$ merge into a single line at $\Fo$, and the cancellation point can be located without calibrated sideband amplitudes.
At cancellation there is no level crossing or gap closing: the smaller spacing $\Fo-2|A|$ reaches its maximum $\Fo$, making the WS spectrum as uniform as possible. 
Quantifying $A$ is therefore required for unbiased quantitative WS spectroscopy and for precision force sensing and interferometry.

The offset $A$ remains only partially understood. 
Tractable analytical results are currently restricted to first order in the strong-field expansion~\cite{MBK15}. 
Floquet–Bloch theory incorporates interband coupling nonperturbatively for periodic multiband lattices~\cite{DeBeule24} and has been applied to the Rice--Mele chain. 
Covariance arguments for bulk systems with conserved particle number and a translationally symmetric field-free Hamiltonian guarantee a multiset of WS ladders with a common real spacing~\cite{ZhangSong25}, but leave the individual offsets undetermined.
To our knowledge, an offset determined exactly within the two-band Floquet description---together with its kinematic bound, zero family, convergence properties, finite-chain consequences, and null readout protocol---remains to be developed.

We address these gaps analytically and numerically, and add experimental readout protocols.
Following the competition between bond and onsite modulations beyond first order and driving it to zero by tuning \(\Delta\) at fixed \(\dt\), we locate the cancellation point at seventh order with an
internal truncation error of \(0.010\Hz\), against \(27.3\Hz\) at first order---a benchmark distinct from the calibration-limited design target of Sec.~\ref{sec:error}. 
The Floquet construction, exact within the two-band model, gives \(|A|\le\Fo/2\) on the principal branch, locates the real zeros of \(A\), and sets the convergence radius through exceptional points at complex detunings; no such singularity intersects the real axis, so the spectrum remains smooth across cancellation. 
For finite chains, a hidden-symmetry defect produces a pairing-violation plateau of height \(2|A|\) (to first order in the modulations and numerically beyond), allowing the offset to be read directly from static spectra without Floquet propagation or time evolution; together with bulk sideband merging, this provides an alternative spectral route to cancellation. 
We combine these signatures with a three-tier tilt-calibration budget and a separate detuning-setpoint uncertainty.

Our  results complement existing work in several ways. 
Exceptional points have been characterized in the resonance spectra of open, continuous bichromatic WS systems~\cite{Elsen11}; in the present two-band problem, complex singularities instead set the convergence of the strong-field expansion while leaving the real-axis spectrum smooth. 
WS sideband spectroscopy~\cite{Kling10} is the closest experimental precedent, but it targets the splitting rather than its cancellation; generalized BOs under linear and quadratic gradients~\cite{Gao26GBO} and winding-number diagnostics~\cite{LeePark15} concern different observables; and two staggered WS ladders are already known~\cite{Breid06,Breid07,Dreisow09}. 
The contribution is therefore not the ladders themselves but the determination of their offset. 
Our analysis is restricted to the single-particle, one-dimensional tight-binding regime.

The paper is organized as follows. 
Section~\ref{sec:model} defines the model and competition mechanism. 
Section~\ref{sec:cancel} develops the analytic-series and Floquet routes to the cancellation point, unifies them in the maximum-uniformity picture, and maps the zero family.
Sections~\ref{sec:finite} and~\ref{sec:obs} present the finite-chain spectral plateau and the experimental signatures and readout protocols, respectively. 
Section~\ref{sec:scope} discusses scope and robustness, and Section~\ref{sec:concl} concludes. Appendices contain proofs, recursions, and numerical prescriptions.

\begin{figure}[tb]
\includegraphics[trim=8 12 15 0, width=\columnwidth]{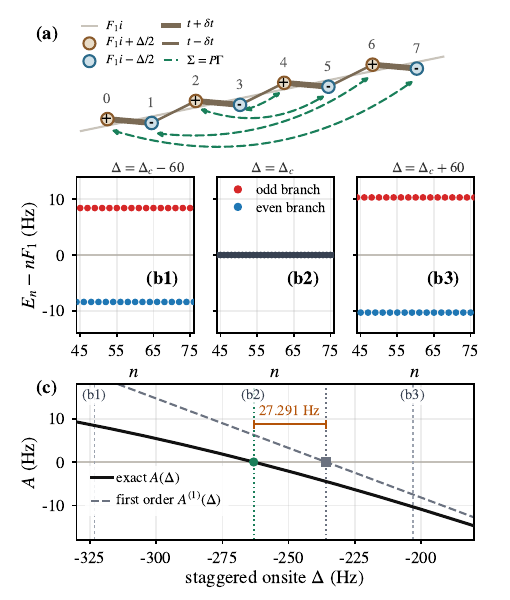}
\caption{
\label{fig:fig1}
Model and competition mechanism.
(a) Tilted Rice--Mele chain with alternating onsite energies \(\pm\Delta/2\), alternating bonds \(t\pm\dt\), and tilt \(\Fo n\). Dashed green arcs: hidden-symmetry pairing \(\Sigma=P\Gamma\).
(b1)--(b3) Bulk spectra \(E_n-n\Fo\) for an even chain with \(N=120\) at \(\Delta=\Dc-60\)~Hz, \(\Dc\), and \(\Dc+60\)~Hz.
Red and blue dots: odd and even branches, respectively. 
At \(\Delta=\Dc\) the two branches are equally spaced, i.e., \(A=0\). 
(c) Staggered amplitude \(A\) versus \(\Delta\). Black solid line: exact \(A(\Delta)\); gray dashed line: first-order \(A^{(1)}(\Delta)\) of Eq.~\eqref{eq:A1}. Vertical lines: detunings of (b1)--(b3).}
\end{figure}

\section{Model and the competition mechanism}
\label{sec:model}

In this section we introduce the minimal driven lattice model and identify the two competing contributions to the staggered amplitude \(A\). 
The Hamiltonian of the tilted Rice--Mele chain on \(N\) sites is
\bea
H&=&-\sum_{n}\bigl[t+(-1)^n\dt\bigr]\bigl(\ket{n}\bra{n+1}+\hc\bigr)
\nonumber\\
&&+\frac{\Delta}{2}\sum_n(-1)^n\ket{n}\bra{n}
+\Fo\sum_n n\,\ket{n}\bra{n} ,
\label{eq:H}
\eea
with \(\Fo\) the tilt per site. Energies are expressed as frequencies throughout. 
The first term is the nearest-neighbor hopping, dimerized with amplitude \(\dt\) about the mean \(t\); the second is a staggered onsite potential of strength \(\Delta\); the third drives the BOs. 
In a double-well interpretation, \(\Delta\) offsets the two wells, while \(\dt\) imbalances the two tunnel couplings. 
All analytic results below hold for general \(\Fo\), \(t\), \(\dt\), and \(\Delta\); numerical evaluations use the representative parameters of Table~\ref{tab:param}, fixed self-consistently in Appendix~\ref{app:num} rather than taken from a specific experiment.

\begin{table}[h]
\caption{\label{tab:param} Representative working point for the numerical
calculations. Energies are in Hz.}
\begin{ruledtabular}
\begin{tabular}{lrlr}
$\Fo$ (tilt)      & $568.0$   & $t$ (mean hopping)   & $226.597$\\
$\Delta$ (onsite) & variable  & $\dt$ (dimerization) & $-47.412$\\
$\dt/t$           & $-0.209$  & $\beta=2t/\Fo$       & $0.797877$\\
$J_0(2\beta)$     & $0.457821$& $J_1(2\beta)$        & $0.569471$\\
$A(0)$            & $-53.9795$ & $2\dt J_1(2\beta)$   & $-53.9995$
\end{tabular}
\end{ruledtabular}
\end{table}

In the bulk, the unmodulated tilted chain is solved exactly by the WS states, Bessel-weighted superpositions \(\bracket{n}{\psi_m}=J_{n-m}(\beta)\) of sites about the center \(m\). 
The tilt is diagonal in this basis. 
Since translation by two sites leaves both modulations invariant and shifts the tilt by \(2\Fo\), the spectrum is invariant under \(E\mapsto E+2\Fo\) and organizes into ladders of spacing \(2\Fo\)---two of them, one per site of the doubled cell. 
After removing the constant part, the most general spectrum of this structure is the ladder of Eq.~\eqref{eq:ladder}, with \(A\) the one parameter left free by the symmetry [Fig.~\ref{fig:fig1}(c)].
The even-chain realization in Fig.~\ref{fig:fig1} serves only to display the bulk ladder cleanly; the odd-chain counterpart is treated in Sec.~\ref{sec:finite} and Fig.~\ref{fig:fig5}.

For finite chains, the hidden symmetry \(\Sigma=P\Gamma\) further enforces the pairing \(E_n+E_{N-1-n}=(N-1)\Fo\) for \(n=0,\ldots,N-1\)---equivalently \(E_n+E_{O-n}=\Fo O\) with \(O=N-1\). 
Exact diagonalization for \(N=40,120,400\) gives \(\max_n|E_n+E_{O-n}-\Fo O|\) at the numerical-noise level, whereas odd \(N\) is treated separately through the defect norm and pairing plateau of Sec.~\ref{sec:finite}. 
The staggering is a bulk property: a WS orbital carries \(98.8\%\) of its weight on the three sites \(m,m\pm1\), with root-mean-square extent \(\beta/\sqrt{2}=0.56\) sites, so \(A\) is independent of \(N\) once the chain exceeds a few tens of sites.

The amplitude \(A\) receives a contribution from each modulation. 
Both enter the same diagonal WS matrix element, with the same alternating phase.
Writing \(V\) for the two modulation terms of Eq.~\eqref{eq:H}, the staggered onsite potential samples the on-site density of a WS orbital, \(\sum_n(-1)^nJ_{n-m}^2(\beta)=(-1)^mJ_0(2\beta)\), and the dimerized hopping its bond density, \(\sum_n(-1)^nJ_{n-m}(\beta)J_{n+1-m}(\beta)=(-1)^mJ_1(2\beta)\), with the two orderings \(\ket{n}\bra{n+1}\) and \(\ket{n+1}\bra{n}\) contributing equally; hence
\be
\Braket{\psi_m}{V}{\psi_m}=(-1)^m\Bigl[\frac{\Delta}{2}J_0(2\beta)
-2\dt J_1(2\beta)\Bigr] ,
\label{eq:Vdiag}
\ee
the \(s=0\) case of Eq.~\eqref{eq:wsme}. 
Comparing with Eq.~\eqref{eq:ladder} identifies the first-order contribution to \(A\) as minus the bracket, giving Eq.~\eqref{eq:A1}. 
The two terms have opposite signs precisely when \(\mathrm{sign}\,\Delta=\mathrm{sign}\,\dt\), both
\(J_0\) and \(J_1\) being positive at the working point (Table~\ref{tab:param})---which is why they compete there. 
The off-diagonal elements \(s\ne0\) of Eq.~\eqref{eq:wsme} generate the higher orders resumed in Sec.~\ref{sec:cancel}.

The diagonal element \eqref{eq:Vdiag} gives the amplitude to first order in the modulations,
\be
A^{(1)}=-\frac{\Delta}{2}J_0(2\beta)+2\dt J_1(2\beta) ,
\label{eq:A1}
\ee
which, after identifying the staggered onsite amplitude of Ref.~\cite{MBK15} with \(\Delta/2\) and the hopping difference with \(-2\dt\), coincides with the first-order strong-field result for a general double-periodic lattice. 
The argument doubles from the \(\beta\) of the WS states to \(2\beta\) in matrix elements---the Neumann addition theorem of Appendix~\ref{app:ws}, already used in Eq.~\eqref{eq:Vdiag}. 
Setting
\(A^{(1)}=0\) gives the first-order cancellation point
\be
\Dc^{(1)}=\frac{4\dt J_1(2\beta)}{J_0(2\beta)} .
\label{eq:Dc1}
\ee
Near cancellation, the two first-order contributions nearly cancel, so higher-order terms that are negligible away from cancellation become important in fixing the cancellation point. 
Their resummation and the resulting convergence radius are determined in Sec.~\ref{sec:cancel}.

\section{The cancellation point: location, nature, and family}
\label{sec:cancel}

Building on the first-order cancellation point of Eq.~\eqref{eq:Dc1}, we determine the cancellation structure in three steps. 
In Sec.~\ref{sec:wsbasis}, an analytic strong-field series in the WS basis locates $\Delta_c$ to $0.01$ ~Hz. 
In Sec.~\ref{sec:floquet}, a nonperturbative two-band Floquet expression supplies the exact zero, its kinematic bound, and its physical interpretation. 
In Sec.~\ref{sec:family}, an analysis of the zero family fixes the convergence radius and the reachability of the remote zeros.

\subsection{Analytic series in the Wannier--Stark basis}
\label{sec:wsbasis}

The goal of this subsection is to turn the competition mechanism into a controlled strong-field expansion of $A$ in powers of $\Delta$ and $\dt$. The starting point is the exact WS matrix element
\bea
\Braket{\psi_m}{H}{\psi_{m+s}}&=&m\Fo\,\delta_{s0}
+\frac{1}{2}(-1)^{m+s}G_s ,\nonumber\\
G_s&=&\Delta J_s(2\beta)+4\dt J_s'(2\beta) ,
\label{eq:wsme}
\eea
with $J_s'$ the derivative with respect to the argument (Appendix~\ref{app:ws}). 
This closed form permits a Rayleigh--Schr\"odinger recursion to any order (Appendix~\ref{app:series}). 
By the joint parity $A(-\Delta,-\dt)=-A(\Delta,\dt)$, the strong-field series contains only odd orders, so the first correction beyond first order is third order; numerically the even-order coefficients vanish at the floating-point floor ($\lesssim10^{-15}$~Hz). 
The result is
\begin{equation}
    \begin{aligned}
        A&=\sum_{m\ \rm odd}\sum_{k=0}^{m}c_{mk}\,\Delta^{\,m-k}\dt^{\,k},\\
        c_{10}&=-\tfrac12J_0(2\beta),\\ 
        c_{11}&=2J_1(2\beta),
    \end{aligned}
    \label{eq:series}
\end{equation}
whose first two coefficients reproduce Eq.~\eqref{eq:A1}.

\begin{figure}[tb]
\includegraphics[trim=5 7 5 0,width=\columnwidth]{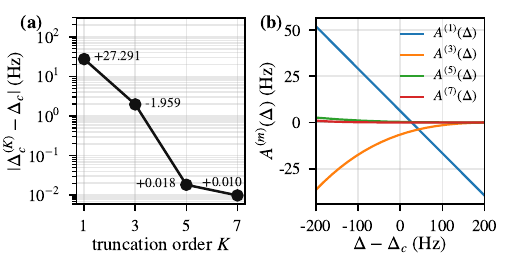}
\caption{
\label{fig:fig2} 
Convergence of the strong-field series for the bulk staggered amplitude $A(\Delta)$.
(a) Cancellation-point error chain $|\Dc^{(K)}-\Dc|$ for $K=1,3,5,7$.
(b) Individual odd-order contributions $A^{(m)}(\Delta)$ near the working point; even orders are parity-forbidden.}
\end{figure}

Truncating Eq.~\eqref{eq:series} at first, third, fifth and seventh order locates \(\Dc\) with the accuracy shown in Fig.~\ref{fig:fig2}(a); the expansion coefficients are fixed by Eq.~\eqref{eq:wsme}, with no free parameters. 
The error falls from \(27.3\)~Hz to \(0.010\)~Hz, a factor of about \(2.7\times10^3\). The cancellation-point error chain is shown in Fig.~\ref{fig:fig2}(a), and the individual odd-order contributions in Fig.~\ref{fig:fig2}(b). 
The series converges inside a convergence disk whose radius \(R\) is determined by its complex-plane singularities in Sec.~\ref{sec:family}; its detailed domain of convergence is analyzed in Sec.~\ref{sec:scope}. 
Generating scripts for all numerical results are listed in Appendix~\ref{app:num} and provided as machine-readable supplementary material.

The convergence test at the working point, whose parameters were fixed self-consistently through the exact value of $\Dc$ itself (Table~\ref{tab:param} and Appendix~\ref{app:num}), so the convergence rate seen there could reflect that fixing rather than the series. 
Repeating the test at two parameter sets chosen independently of the exact zero removes the ambiguity: the sub-$0.01$~Hz seventh-order precision is reproduced there, and is therefore a property of the series itself (Table~\ref{tab:ind}, Appendix~\ref{app:num}).

The series locates $\Dc$ to $0.010$~Hz, but it remains a perturbative construction: the exact zero, its kinematic bound, and its relation to the zero family and convergence radius all call for a non-perturbative expression.

\begin{figure}[tb]
\includegraphics[trim=10 8 5 0, width=\columnwidth]{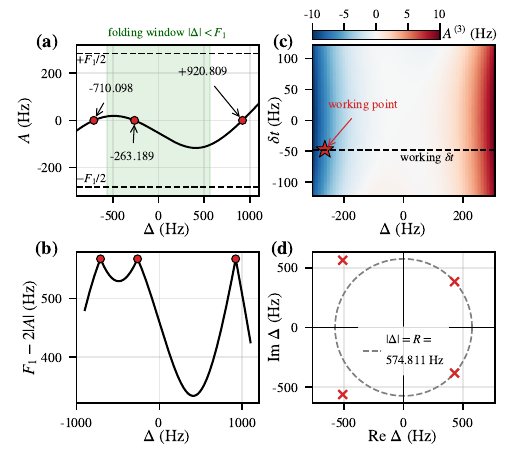}
\caption{
\label{fig:fig3} 
Global structure of the bulk staggered amplitude $A(\Delta)$.
(a) Real-axis $A(\Delta)$ from Eq.~\eqref{eq:Afloq}. 
Green shading: folding window $|\Delta|<\Fo$. Dashed lines: kinematic bound $\pm\Fo/2$. Red points: zeros $\Dcm$, $\Dc$, $\Dcp$.
(b) Minimum level spacing $\Fo-2|A|$ versus $\Delta$; the maximum $\Fo$ is attained at $\Dc$. 
(c) Third-order correction $A^{(3)}(\Delta,\dt)$ in the $(\Delta,\dt)$ plane. Star: working point.
(d) Complex-plane singularities of $A$: two conjugate pairs of exceptional points of $U$ (crosses), with $|\Delta_{\rm EP}|= 574.811$~Hz (circle) setting the radius of convergence.}
\end{figure}

\subsection{Floquet propagator and the maximum-uniformity picture}
\label{sec:floquet}

A two-band Floquet propagator over one Bloch period of the doubled cell gives the amplitude nonperturbatively and, within that model, exactly. 
It complements the WS series of Sec.~\ref{sec:wsbasis}: the series is controlled near the working point, but its radius is set by exceptional points, whereas Eq.~\eqref{eq:Afloq} defines the amplitude globally within the two-band model. 
In this construction, the double-cell Hamiltonian \(h(k)\) is driven along \(k(t)=k_0-4\pi\Fo t\), and the trace of its propagator \(U\) over one period \(T=1/(2\Fo)\) gives
\be
A=\frac{\Fo}{\pi}\arcsin\!\Bigl(-\frac{i}{2}\mathrm{Tr}\,U\Bigr) .
\label{eq:Afloq}
\ee
We take the principal branch of the arcsine throughout, which folds the quasi-energy difference into the Brillouin zone \([-\Fo/2,\Fo/2]\); for real \(\Delta\) the argument of the arcsine then lies in \([-1,1]\) and is independent of the initial phase \(k_0\). 
Equation~\eqref{eq:Afloq} is exact within the two-band model and holds uniformly in the field strength.
Its construction and numerical verification are given in Appendix~\ref{app:floquet}, together with a discussion of why the naive prescription of driving the Rice--Mele modulation alone fails. 
The propagator \(U\) itself is constructed numerically, by fourth-order Magnus integration; Eq.~\eqref{eq:Afloq} is thus an exact relation within the two-band model, i.e., not a parameterized analytic function of the Hamiltonian parameters.

Equation~\eqref{eq:Afloq} yields a kinematic bound. 
It is the symmetry of Sec.~\ref{sec:model} expressed in the folded language: the two subladders
are defined only modulo the \(2\Fo\) of a two-site translation, so their offset \(2A\), and hence \(A\) itself, is defined only modulo \(\Fo\). 
The representative selected by the principal branch obeys
\be
|A|\le\frac{\Fo}{2} .
\label{eq:bound}
\ee
The bound gives the cancellation point a precise meaning. 
Adjacent spacings follow from Eq.~\eqref{eq:ladder},
\be
E_{n+1}-E_n=\Fo+2(-1)^{n}A ,
\label{eq:spacings}
\ee
so the minimum spacing is
\be
\min_n(E_{n+1}-E_n)=\Fo-2|A| .
\label{eq:minspacing}
\ee
At \(A=0\), this minimum attains its largest value, \(\Fo\) [Fig.~\ref{fig:fig3}(b)]: the cancellation point is the point of maximum spectral uniformity, not a level crossing or a phase transition. 
The levels remain non-degenerate, with uniform internal spacing \(\Fo\), and only the beat period \(\tau=1/(2|A|)\) diverges, while the spectrum itself remains regular.

It is also not a Rice--Mele gap closing, which occurs only at \(\Delta=\dt=0\).\footnote{At \(\Dc\), the untilted Rice--Mele gap is \(\sqrt{\Delta^{2}+16\dt^{2}}=324.4\)~Hz.} 
Chiral symmetry, which would make such a closing topologically significant, would require \(\Delta=\Fo=0\) and is absent here: the tilt localizes the eigenstates into WS orbitals and replaces the gap-closing criterion with one of spectral uniformity. 
Accordingly, \(\Dc\) is a spectral property of the tilted chain, with no topological classification attached in this setting.
The third-order correction landscape around the working point is shown in Fig.~\ref{fig:fig3}(c). 
The restoration of a uniform ladder at \(\Dc\) is shown directly in Fig.~\ref{fig:fig1}(b1--b3).

At the working parameters, \(\Dc\) is the unique zero of \(A(\Delta)\) inside the folding window \(|\Delta|<\Fo\); the full zero family, including the remote zeros, is analyzed in Sec.~\ref{sec:family}.

\begin{figure}[tb]
\includegraphics[trim=10 10 2 0, width=\columnwidth]{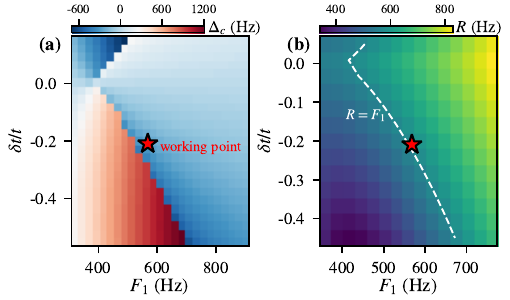}
\caption{\label{fig:fig4} 
(a) Nearest zero \(\Dc(\Fo,\dt/t)\) in the \((\Fo,\dt/t)\) plane on a \(25\times25\) grid. Color encodes \(\Dc\). 
(b) Nearest exceptional-point radius \(R=|\Delta_{\rm EP}|\) on a \(15\times13\) grid. 
Color encodes \(R\). 
The map is obtained by continuity following of both \(z=\pm1\) branches and taking the smaller radius at each point. Dashed contour: \(R=\Fo=568\)~Hz.}
\end{figure}

\begin{figure*}[tb]
\includegraphics[width=\textwidth]{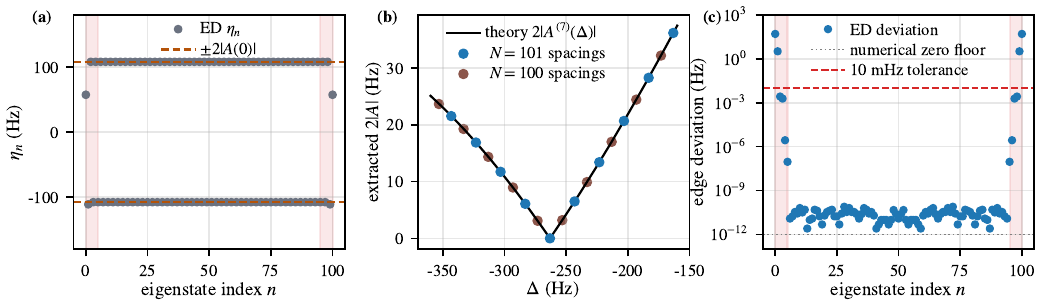}
\caption{\label{fig:fig5} 
Finite-chain signatures of the staggered amplitude.
(a) Pairing residual \(\eta_n=E_n+E_{O-n}-\Fo O\) for an odd chain with \(N=101\), \(O=N-1\), at \(\Delta=0\). Gray dots: ED \(\eta_n\). Orange dashed lines: \(\pm 2|A(0)|=\pm 107.959\)~Hz. 
Pink shaded regions mark the edges.
(b) Extracted \(2|A(\Delta)|\) versus \(\Delta\) from the alternating part of nearest-neighbor spacings. Blue dots: odd chain \(N=101\); brown dots: even chain \(N=100\); black solid line: theoretical strong-field series \(2|A^{(7)}(\Delta)|\). 
(c) Edge deviation \(\bigl|\,|\eta_n|-2|A(0)|\,\bigr|\) versus eigenstate index \(n\) on a logarithmic scale, for the same parameters as (a). 
Blue dots: ED deviation; dotted gray line: numerical zero floor; red dashed line: \(10\)~mHz tolerance. }
\end{figure*}

\subsection{Zero family and analytic structure}
\label{sec:family}

The zeros of \(A\) are physical spectral features: they mark the restoration of spectral equidistance on the real axis. 
The analytic structure of \(A\) is a separate mathematical property: the exceptional points in the complex plane determine the radius of convergence and which zeros are reachable by the series. 
Two distinct criteria enter here: the folding window \(|\Delta|<\Fo=568\)~Hz, set by quasi-energy folding and containing the physical working point, and the convergence disk \(|\Delta|<R\), set by the complex-plane singularities and limiting the series. 
At our parameters the two scales differ by only \(1\%\), but they are conceptually distinct.

The Floquet construction shows that \(A(\Delta)\) oscillates along the real axis and vanishes at more than one detuning [Figs.~\ref{fig:fig3}(a) and~\ref{fig:fig4}(a)]. 
At our parameters, three zeros lie within \(|\Delta|\le2.5\Fo\):
\begin{equation}
    \begin{aligned}
        &\Dcm=-710.0978\Hz ,\\
        &\Dc=-263.1889\Hz ,\\
        &\Dcp=+920.8092\Hz .
    \end{aligned}
\end{equation}
Between the outer zeros (\(|\Delta|\lesssim900\)~Hz), the largest value is \(|A|=117.218\)~Hz at \(\Delta=408.6\)~Hz, well below the kinematic bound \(\Fo/2=284\)~Hz. 
All three are bulk properties and are stable with increasing \(N\), with edge-state exclusion and numerical tests detailed in Appendix~\ref{app:num}.

Physically, the remote zeros realize the same spectral mechanism as the working point---restoration of spectral equidistance, \(A=0\)---at larger detunings. 
They are genuine predictions of the two-band model, although observing them would require \(|\Delta|\gtrsim700\)~Hz, at the edge of the deep-well regime, where the model itself must be verified. 
They do not interfere with the working point: the zero of \(A(\Delta)\) at \(\Dc\) is simple, with \(|dA/d\Delta|\simeq0.157\), so the merging condition at \(\Dc\) is locally independent of the remote zeros.

Only \(\Dc\) lies inside the folding window; the two remote zeros lie outside the convergence disk of Eq.~\eqref{eq:series}, and the nearer one is reached by resummation rather than by truncation. 
The zeros form branches in the \((\Fo,\dt)\) plane that appear and disappear as the field is varied [Fig.~\ref{fig:fig4}(a)]; the branch structure changes character where \(J_0(4t/\Fo)=0\), i.e., at \(\Fo=376.9\)~Hz for our hopping (Appendix~\ref{app:sing}).

Fig.~\ref{fig:fig3}(d) locates the singularities that fix the radius of convergence of the series. \(A(\Delta)\) is singular where the two eigenvalues of \(U\) coalesce, i.e., where \(\mathrm{Tr}\,U=\pm2i\); this is an exceptional point in the usual sense~\cite{Heiss12}. 
At such a point \(A\) loses analyticity as a function of \(\Delta\), and the nearest such singularity sets the Taylor radius. Solving for \(\mathrm{Tr}\,U=\pm2i\) gives two conjugate pairs, \(\Delta_{\rm EP}=428.075\pm383.614i\)~Hz and \(-510.326\pm562.712i\)~Hz, so the Taylor series about the origin has radius of convergence
\be
R=|\Delta_{\rm EP}|=574.811\Hz .
\label{eq:R}
\ee
Nothing singular happens on the real axis. The consequence is visible in Fig.~\ref{fig:fig3}(d): \(\Dc=-263.1889\)~Hz lies well inside the disk, whereas the two remote zeros lie outside it, which is why no sequence of Taylor truncations converges at them.

A Pad\'e approximation extends the domain of validity on the real axis.
Despite the radius \(R\), it also reaches the nearer remote zero: \(\Dcm\) is reproduced to the \(0.1\)-Hz level (\([10/10]\): \(0.12\)~Hz; \([19/6]\): \(10^{-3}\)~Hz), whereas \(\Dcp\) is recovered only erratically, from a few Hz to several hundred Hz, without convergence in the order (Appendix~\ref{app:sing}). 
Resummation thus recovers part of what no convergent Taylor sequence reaches, but not all of it.

\section{Finite chains: hidden symmetry, spectral inversion, and the odd-\(N\) plateau}
\label{sec:finite}

For even chains, the hidden symmetry gives an exact spectral pairing that contains no information about the staggered offset \(A\). 
For odd chains, the breaking of that symmetry produces a pairing plateau of height \(2|A|\) that persists beyond first order, as verified numerically, making \(A\) directly readable from the static spectrum.

\subsection{Even chains: hidden symmetry and spectral inversion}

The staggering of Sec.~\ref{sec:model} is a bulk property, fixed there by two-site translation. On a finite chain there is no translation, but for even \(N\) the spectrum satisfies an exact pairing. With the staggered sign \(\Gamma\ket{n}=(-1)^n\ket{n}\) and reflection \(P\ket{n}=\ket{O-n}\), where \(O=N-1\), the composite operator
\be
\Sigma=P\Gamma
\label{eq:Sigma}
\ee
acts as a spectral inversion:
\be
\Sigma H\Sigma^{-1}=\Fo O-H ,
\label{eq:specinv}
\ee
so that
\be
\Sigma\ket{E_n}=\ket{\Fo O-E_n}.
\label{eq:pairing}
\ee
Equation~\eqref{eq:pairing} is exact for every even \(N\), not a large-\(N\) statement. 
Like chiral and Floquet spectral symmetries, \(\Sigma\) inverts rather than commutes with the Hamiltonian~\cite{Asboth16,Kitagawa10,Rudner13}; what is new here is its spectral consequences in the WS problem, including the odd-chain plateau and its use in Sec.~\ref{sec:obs}.

We define the pairing residual
\be
\eta_n=E_n+E_{O-n}-\Fo O ,
\label{eq:eta}
\ee
which measures the failure of Eq.~\eqref{eq:pairing}. 
By Eq.~\eqref{eq:pairing}, \(\eta_n=0\) exactly on every even chain. 
The exact pairing fixes the inversion center of the spectrum but leaves the alternating part unconstrained: the even chain is blind to \(A\). 
Hence \(|A|\) must be read from the alternating spacings,
\be
|A|=\frac{|E_{n+1}-E_n-\Fo|}{2} ,
\label{eq:Afromspacing}
\ee
rather than from the pairing; the uniform spacing \(\Fo\) must be subtracted before the staggered offset is extracted.

On an odd chain, the central site has no reflection partner, so Eq.~\eqref{eq:pairing} is lost. The residual no longer vanishes. 
Exact diagonalization shows that \(\eta_n\) does not vanish as \(N\) increases away from cancellation, but saturates at a nonzero plateau [Fig.~\ref{fig:fig5}(a--c)]. 
That plateau is the finite-size spectral signature of the staggered amplitude \(A\).

\subsection{Odd chains: defect norm and the pairing plateau}
\label{sec:odd}

The breaking of \(\Sigma\) on an odd chain is controlled by a bulk symmetry-breaking operator carried by every site and every bond, not by a spatial lattice defect or an edge state. 
Write
\be
H=H_0+V ,
\label{eq:HV}
\ee
where \(H_0\) is invariant under \(\Sigma\) and
\be
V=\frac12 D,\qquad D=2(H_{\dt}+H_{\rm stag}) ,
\label{eq:defect}
\ee
with \(H_{\dt}\) and \(H_{\rm stag}\) the hopping and staggered-onsite terms of Eq.~\eqref{eq:H}, with their signs as written in Eq.~\eqref{eq:H}. 
The defect \(D\) anticommutes with \(\Sigma\) and is odd under the joint inversion \((\Delta,\dt)\to(-\Delta,-\dt)\); it vanishes only when both modulations do. 
Since \(D\) is tridiagonal, with both diagonal and off-diagonal entries, its spectral norm follows from the square of a single two-by-two cell block,
\be
\|D\|=\sqrt{\Delta^2+16\dt^2\cos^2\frac{\pi}{N+1}} ,
\label{eq:defectnorm}
\ee
where the cosine factor is the norm of the signed adjacency of the open chain (Appendix~\ref{app:odd}). 
The defect is carried by every site and every bond, and its norm approaches a chain-independent value as \(N\) grows; the breaking is therefore a parity effect, not a surface phenomenon.

The norm vanishes with the modulations and controls the pairing violation through the Weyl bound
\be
|\eta_n|\le\|D\| ,
\label{eq:oddbound}
\ee
for every \(N\) and \(\Delta\). 
Away from cancellation the bound is \(\sqrt{\Delta^2+16\dt^2}\) to leading order in \(1/N\); at \(\Dc\), the bound equals \(\sqrt{\Dc^2+16\dt^2}\) to the same order.

The bound explains why the residual must remain nonzero, but not its plateau height. 
Away from the edges, the eigenstates are bulk-like and can be paired according to the bulk form of Eq.~\eqref{eq:ladder}, excluding the edge states; projecting onto these pairs gives the plateau height.
Expanding the defect expectation value to first order in \(V\) and projecting onto the paired subspace gives
\be
\eta_n\simeq 2(-1)^{n+1}A
\qquad\text{(odd \(N\), bulk pairs)},
\label{eq:plateau}
\ee
so the plateau magnitude is \(2|A|\), not \(\|D\|\) itself. 
The sign alternates with the pair; in practice one uses the plateau magnitude. 
The plateau is thus a direct finite-chain signature of the staggered amplitude. 
A proof of Eq.~\eqref{eq:plateau} to first order, together with its numerical verification beyond first order, is given in Appendix~\ref{app:odd}; the statement is verified numerically rather than
elevated to a theorem.

For odd chains, with the levels near the edges excluded, the plateau height extracted from \(\eta_n\) agrees with \(2|A|\) to numerical accuracy throughout the working region and does not drift with \(N\) for \(N=51\)--\(401\) (the extracted plateau agrees with \(2|A(0)|\) within \(10^{-10}\)~Hz); the numerical protocol is given in Appendix~\ref{app:odd}. 
The same bulk amplitude is obtained from the alternating part of nearest-neighbor spacings, as shown in
Fig.~\ref{fig:fig5}(b). 
The result is not an edge effect. The leading edge correction to the plateau is bounded by
\be
|\eta_{\rm edge}|\le4t\,J_d\!\left(\frac{2t}{\Fo}\right)^2 ,
\label{eq:edgebound}
\ee
where \(d\) is the distance to the nearer edge and \(J_d\) is the corresponding Bessel tail. Equation~\eqref{eq:edgebound} gives \(N\gtrsim16\) for a \(10\)~mHz tolerance at the parameters of
Table~\ref{tab:param}; the derivation is given in Appendix~\ref{app:odd}.

Two consequences make the plateau useful. 
First, at cancellation the plateau is zero and the residual is dominated by the edge correction alone, so the plateau zero identifies \(\Dc\) on the finite chain. 
Second, at fixed \(\Fo\) and \(\dt\), locating that zero within the folding window \(|\Delta|<\Fo\) determines \(\Dc\) without the WS basis, without Floquet propagation, and without time evolution---the offset is read directly from the static spectrum, and the value agrees with the bulk methods of
Secs.~\ref{sec:wsbasis}--\ref{sec:family}. Section~\ref{sec:obs} turns this plateau, together with the bulk sideband merging, into readout protocols and an error budget.

\begin{figure}[tb]
\includegraphics[width=\columnwidth]{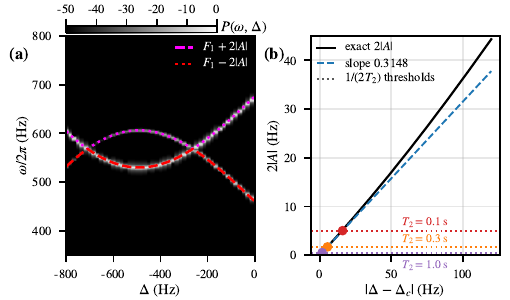}
\caption{\label{fig:fig6} 
Beat spectroscopy.
(a) Log-power spectrum \(P(\omega,\Delta)\) of the sublattice imbalance versus \(\Delta\) and \(\omega/2\pi\) for \(N=60\), evolution time \(1.0\)~s, and lower-sublattice fraction \(2.0\%\). Magenta and red dashed lines: \(\Fo+2|A|\) and \(\Fo-2|A|\), respectively.
(b) Onset of \(2|A|\) versus \(|\Delta-\Dc|\). Black solid line: exact \(2|A|\); blue dashed line: linear slope \(0.3148\). Horizontal dotted lines: \(1/(2T_2)\) thresholds for \(T_2=0.1\)~s (red), \(0.3\)~s (orange), and \(1.0\)~s (purple).}
\end{figure}

\section{Experimental signatures and protocols}
\label{sec:obs}

This section turns the preceding theoretical results into experimental signatures, two readout protocols, and a calibration-aware error budget.

\subsection{Time-domain and spectral signatures}

Prepare a bulk-centered wave packet on a single sublattice, with a width comparable to the WS localization length (rms \(\beta/\sqrt{2}=0.56\) sites, Sec.~\ref{sec:model}), and monitor the sublattice imbalance; the preparation and coherence requirements are detailed in Appendix~\ref{app:protocols}. 
With the response band-passed about \(\Fo\) to suppress higher harmonics, the real imbalance takes the leading form
\be
S(t)\propto \cos(2\pi\Fo t)\cos\bigl(2\pi(2A)t\bigr) ,
\label{eq:beat}
\ee
where \(S(t)=\langle\psi(t)|\hat I|\psi(t)\rangle\) and \(\hat I=\sum_j(-1)^j\hat n_j\). The spectrum shown in Fig.~\ref{fig:fig6}(a) is the normalized log-power
\begin{equation}
    \begin{aligned}
        & P(\omega,\Delta)=10\log_{10}\frac{|\widetilde S(\omega,\Delta)|^2} {\max_{\omega}|\widetilde S(\omega,\Delta)|^2}, \\
    & \widetilde S(\omega,\Delta)=\int_0^{T_{\rm ev}} [S(t)-\overline S]\, w(t)e^{i\omega t}\,dt ,
    \end{aligned}
    \label{eq:imbalance-spectrum}
\end{equation}
with window function \(w(t)\) and evolution time \(T_{\rm ev}\). 
The product-to-sum spectrum of Eq.~\eqref{eq:beat} consists of two sidebands at \(\Fo\pm2|A|\). Thus the sideband splitting is \(4|A|\), while the envelope beat has frequency \(2|A|\) and period
\be
\tau=\frac{1}{2|A|} .
\label{eq:tau}
\ee

At \(\Delta=0\), \(|A|=53.9795\)~Hz and \(\tau=9.263\)~ms, well within reach of current coherence times; the sidebands then sit at \(460.0\) and \(676.0\)~Hz [Fig.~\ref{fig:fig6}(a)]. 
The envelope discrimination is evident from the spectrum. A full-wave rectification of the envelope doubles the apparent beat frequency to \(4|A|\), which can simplify the time-domain extraction.  
For the packet used here, the principal out-of-band contribution is a harmonic at \(2\Fo\) at \(37\%\) of the stronger sideband, while a line at \(3\Fo\) and the lower-sublattice fraction remain below \(6.4\%\); the bandpass suppresses these residuals.
Related St\"uckelberg interferometric readouts of Bloch oscillations have been demonstrated in accelerated lattices~\cite{Kling10}.

At \(\Dc\), the two sidebands merge into a single line at \(\Fo\): the doublet becomes a singlet and the beat disappears. 
This is the primary experimental signature. It is a null measurement requiring no calibrated sideband amplitudes, although the tilt and line frequency still need to be calibrated.

\subsection{Readout protocols}

\emph{Protocol A (consistency map).} 
Equation~\eqref{eq:Afloq} is exact within the two-band model and, once the working point is fixed, contains no additional adjustable parameters. 
The whole curve \(A(\Delta)\)---its three zeros, its window maximum \(|A|=117.218\)~Hz, and the bound \(|A|\le\Fo/2\)---therefore constrains the theory far more tightly than any single number. 
The only inputs are \(\Fo\), \(t\) and \(\dt\). 
The remote zeros, however, require \(|\Delta|\gtrsim700\)~Hz and serve as secondary whole-curve discriminators, while \(\Dc\) remains the primary experimental target.

\emph{Protocol B (sideband amplitude).} 
Near \(\Dc\) the sideband amplitude is linear in \(A\), and hence in \(\Delta-\Dc\), with slope \(|dA/d\Delta|=0.15742\) at \(\Dc\) (the first-order formula would give \(0.2289\), overestimating the slope by \(45\%\); the slope of \(2|A|\) with respect to \(\Delta\) is twice this value). 
One may therefore attempt to locate \(\Dc\) by extrapolating the sideband amplitude to zero. 
The readout itself is linear over the relevant range, but the extrapolation is not: over \(|\Delta-\Dc|\lesssim120\)~Hz, a purely linear extrapolation misses \(\Dc\) by \(\approx 9\)~Hz (\(9.48\)~Hz for a \(20\)~Hz grid, varying between \(8.7\) and \(9.5\)~Hz for \(10\)--\(20\)~Hz steps), with a bias set by the sampling geometry. 
Protocol B is therefore a local simplification of Protocol A, not an independent replacement for it: its advantage is that it requires only a local scan around \(\Dc\), but a shape-constrained fit is still required, as in Protocol A. 
The simulation details are given in Appendix~\ref{app:protocols}.

\subsection{Error budget}
\label{sec:error}

The uncertainty has two conceptually distinct parts: the theoretical prediction of \(\Dc\), propagated from the calibrated parameters \(\Fo\), \(t\) and \(\dt\), and the experimental location of that prediction along the \(\Delta\) axis, limited by the detuning setpoint and readout. 
A conservative worst-case propagation is
\be
\delta\Dc\le\delta\Delta+
\left|\frac{\partial\Dc}{\partial\Fo}\right|\delta\Fo+
\left|\frac{\partial\Dc}{\partial t}\right|\delta\,t+
\left|\frac{\partial\Dc}{\partial\dt}\right|\delta\dt ,
\label{eq:propagation}
\ee
where \(\delta\Delta\) combines the detuning-setpoint uncertainty and the spectral readout resolution, while the remaining terms propagate the theoretical-parameter calibration uncertainty. 
The exact slopes are \(\partial\Dc/\partial\Fo=1.5255\), \(\partial\Dc/\partial t=-3.4900\) and \(\partial\Dc/\partial\dt=7.1465\), from the propagator and diagonalization.

The ranking depends on the measure. Per unit absolute error, the dimerization carries the largest derivative (\(7.15\) versus \(3.49\) and \(1.53\)). 
Per unit relative error, the tilt dominates (\(|\partial\Dc/\partial x|\,|x|=867\), \(791\) and \(339\)~Hz for \(\Fo\), \(t\) and \(\dt\)). 
Under the uncertainties \((\delta\Fo,\delta\,t,\delta\dt)=(16.4,50,20)\)~mHz, the linear worst-case bound gives \(0.025+0.175+0.143=0.342\)~Hz; if the three independent terms are instead combined in quadrature, the nominal estimate is \(0.227\)~Hz. 
Combined, the four contributions place the realistic design target for a first measurement at the \(0.2\)--\(0.3\)~Hz level (\(0.23\)~Hz in quadrature, \(0.34\)~Hz worst case); the \(0.010\)~Hz figure of the analytic series remains a theory-internal benchmark, whose standing is directly supported by the independent-parameter convergence tests of Table~\ref{tab:ind}.

The three-tier budget discussed here propagates the tilt calibration only; it does not include \(\delta\,t\), \(\delta\dt\) or \(\delta\Delta\).
Taking \(\delta\Fo=164\), \(53.8\) and \(16.4\)~mHz for minimal, standard and high-stability field calibration gives \(\delta\Dc=0.2499\), \(0.0821\) and \(0.0250\)~Hz from the tilt term alone. With \(\delta\,t\) and \(\delta\dt\) fixed at \(50\) and \(20\)~mHz, the corresponding quadrature totals are \(0.337\), \(0.240\) and \(0.227\)~Hz. 
Reaching a quadrature-combined uncertainty of \(0.010\)~Hz requires an effective-tilt uncertainty \(\delta F_{1,\rm eff}\lesssim5.7\)~mHz together with \(\delta\Delta\lesssim5\)~mHz, where \(F_{1,\rm eff}\) is the tilt entering the measured WS spacing. 
Indeed, the tilt term contributes \(1.5255\times5.7\)~mHz \(=8.7\)~mHz and the detuning-setpoint term contributes \(5\)~mHz:
\be
\sqrt{(1.5255\times5.7~{\rm mHz})^2+(5~{\rm mHz})^2}=10.0~{\rm mHz},
\ee
so the quoted \(0.010\)~Hz figure is a quadrature design estimate rather than the conservative linear bound of Eq.~\eqref{eq:propagation}. 
A tilt uncertainty of \(6.6\)~mHz would on its own already contribute \(0.0101\)~Hz in this design estimate, so the two requirements are not independent; they amount to millihertz-level self-calibration of both the tilt and the detuning setpoint. 
This is demanding, but it is feasible for a lattice in which the tilt is itself interferometrically referenced~\cite{Cadoret08,Bouchendira11,Genske13}; existing WS lattice-clock platforms, in which individual WS states are already resolved spectroscopically~\cite{Aeppli22}, are natural candidates.

\section{Scope, robustness, and open questions}
\label{sec:scope}

Four issues bound the scope of these results. 
The first is numerical and has been treated above; the other three bear on an experiment.

\emph{Numerical convergence.} 
The all-orders expansion of Sec.~\ref{sec:wsbasis} is a bivariate series in \((\Delta,\dt)\) truncated at total degree \(m\) [Eq.~\eqref{eq:series}]. 
Along the working line it converges inside the disk \(|\Delta|<R\), whose radius is set in Sec.~\ref{sec:family}. 
The full bivariate convergence domain is not characterized analytically, and accuracy away from the working line is established numerically (Tables~\ref{tab:ind} and~\ref{tab:domain}).
Successive orders improve the result up to \(|\Delta|\lesssim320\)~Hz, with the optimal truncation order decreasing farther out---at \(|\Delta|=400\)~Hz the error falls from \(0.48\)~Hz at seventh order to \(2.4\times10^{-3}\)~Hz at fifteenth (eight odd sectors, \(m\le15\)), and is about \(30\) times larger at \(\Delta=+400\)~Hz than at \(-400\)~Hz, consistent with the asymmetric location of the complex exceptional points (Sec.~\ref{sec:family}). 
Whether a systematic analytic continuation can replace the Floquet construction beyond the convergence disk remains open.

\emph{Lattice imperfections and confinement.} 
Next-nearest-neighbor tunneling \(t_2\), typically a few percent of \(t\) in a double-well superlattice, leaves \(\Dc\) unchanged. 
In the WS basis \(\langle\psi_m|H_{t_2}|\psi_{m'}\rangle=-t_2\,\delta_{m',m\pm2}\) has no diagonal or staggered component, because \(\sum_nJ_n(x)J_{n\pm2}(x)\) vanishes identically. 
The deeper reason is exact: in the period-2 cell, \(t_2\) contributes \(-2t_2\cos k\) equally to both sublattices, so its one-period average vanishes. Direct diagonalization confirms this cancellation: for \(-10\%\le t_2/t\le10\%\), \(\Dc\) moves by less than \(3\times10^{-12}\)~Hz.

Harmonic confinement, by contrast, makes the tilt site dependent, \(\Fo\to\Fo(n)=\Fo+\kappa n\), so the measured amplitude samples \(2t/\Fo(n)\). 
Averaged over a Gaussian cloud of rms width \(10\) sites with \(\kappa=0.1\)~Hz/site, the centroid of \(A\) shifts by only \(3\times10^{-4}\)~Hz, while the local displacement of \(\Dc\) produces a
residual floor \(|\partial A/\partial\Delta|\,|\partial\Dc/\partial\Fo|\,\sigma_{\Fo} =0.24\,\sigma_{\Fo}\)~Hz. 
Keeping this floor below \(0.01\)~Hz requires \(\sigma_{\Fo}\lesssim42\)~mHz. Inhomogeneous broadening of the sidebands is less restrictive: their splitting is spread by \(0.96\,\sigma_{\Fo}\), below the \(1/T_2\) linewidth for the coherence times of Sec.~\ref{sec:obs}.

\emph{Band closure and interband corrections.} 
The two-band model requires the deep-lattice regime, in which the Bloch frequency and the modulation
amplitudes are small compared with the higher-band gap: \(\Fo,|\Delta|\ll\Delta_{\rm gap}\). 
Since Protocol~A reaches \(|\Delta|\simeq0.92\)~kHz, an experiment must satisfy \(\Delta_{\rm gap}\gg0.92\)~kHz together with \(\Fo=0.57\)~kHz; the gap itself is set by the lattice depth. 
Interband corrections emphasized in the nonperturbative Floquet--Bloch treatment of Ref.~\cite{DeBeule24} renormalize \(t\) and \(\dt\) smoothly with the lattice depth. 
Such renormalizations are absorbed into the representative values of Table~\ref{tab:param}, or into independently calibrated values in an experiment, while residual Zener-resonance corrections at the Bloch scale lie outside the two-band model. 
We therefore expect the absolute value of \(\Dc\) to be limited by the calibration of \(\Fo\), \(t\) and \(\dt\) (Sec.~\ref{sec:obs}), while the existence of the cancellation point and its shape-constrained location are robust against smooth perturbations of this kind.

\emph{Identifiability and falsifiable outputs.} 
The values of Table~\ref{tab:param} are representative rather than measured: they are fixed self-consistently in Appendix~\ref{app:num} by \(|A(0)|=53.9795\)~Hz, \(\Dc^{(1)}=-235.9\)~Hz and \(\Dc=-263.1889\)~Hz. 
In an experiment, \(\Fo\), \(t\) and \(\dt\) would instead come from independent calibration. The self-consistent fix is well posed: the Jacobian of the three conditions with respect to \((\Fo,t,\dt)\) has singular values \((9.89,0.47,0.07)\) and condition number \(1.5\times10^2\), so all inputs are identifiable, with \(\dt\) the best-determined direction and a mild degeneracy in the \((\Fo,t)\) plane. 
Quantities tied to the representative fix itself---the zeros and the convergence sequence shown in Fig.~\ref{fig:fig2}(a)---test internal consistency rather than providing independent predictions. 
The independently falsifiable outputs are the slopes, the window maximum, the bounds, the plateau values, and the zero positions obtained for independently calibrated parameters.

\section{Conclusions}
\label{sec:concl}

We have determined the staggered WS offset \(A\) of a tilted Rice--Mele chain, and with it the cancellation point \(\Dc\), exactly within the two-band Floquet description. 
In the WS basis the strong-field expansion reduces to Bessel algebra; at the working point of Table~\ref{tab:param}, seventh-order truncation locates \(\Dc=-263.1889\)~Hz with an internal truncation error of \(0.010\)~Hz, against \(27.3\)~Hz at first order. 
A two-band Floquet propagator gives the amplitude exactly within that model, \(A=(\Fo/\pi)\arcsin(-i\,\mathrm{Tr}\,U/2)\), implying the kinematic bound \(|A|\le\Fo/2\). 
Each real zero of \(A\) is a point of maximum spectral uniformity, with minimum spacing attaining \(\Fo\) and no degeneracy or gap closing. 
Three real zeros lie within \(|\Delta|\le2.5\Fo\) at the working point; only \(\Dc\) lies inside the folding window. 
Complex exceptional points set the convergence radius \(R=574.811\)~Hz, placing the remote zeros outside the domain of convergent Taylor truncation; Pad\'e resummation recovers only the nearer one.

The offset can be read from the static spectrum without Floquet propagation or time evolution: on an odd-length chain, the hidden-symmetry defect produces a pairing plateau of height \(2|A|\), while in the bulk the sidebands merge into a single line at \(\Dc\). 
These are the most promising experimental signatures; the remote zeros serve as secondary whole-curve discriminators. 
As in Sec.~\ref{sec:obs}, the \(0.010\)~Hz figure above is a theory-internal benchmark, not the experimental design target, which is \(0.2\)--\(0.3\)~Hz; matching the former experimentally requires the millihertz-level self-calibration of both the effective tilt and the detuning setpoint quantified there.

Four families of independently testable outputs remain. 
The zero position within the folding window probes the Floquet expression and the working-point cancellation; the maximum of \(|A|\) and the kinematic bound probe the principal-branch folding; the slope at cancellation, read out with a shape-constrained fit, probes the all-order shift of \(\Dc\); and the odd-chain plateau values probe the hidden-symmetry defect. 
The analysis is restricted to the single-particle, one-dimensional tight-binding regime, and several questions remain open. The first is interactions: the budget of Sec.~\ref{sec:obs} would resolve an interaction-induced ladder shift only down to the \(10\)-mHz scale, so a larger shift would appear as a
displacement of \(\Dc\). 
Mean-field shifts and interaction-induced dephasing of Bloch oscillations are experimentally documented in tilted lattices~\cite{Morsch01,Gustavsson08}, and Floquet--Bloch treatments of the interacting problem exist~\cite{Kolovsky03,Buchleitner03,Kolovsky10}.
Beyond interactions, open questions include disorder and finite-temperature effects on the odd-chain plateau, higher-dimensional WS ladders, and experimental verification of the remote zeros beyond the deep-well regime, where comparison with Stark-localization phenomenology~\cite{Schulz19,Nieuwenburg19,Taylor20,Guardado20,Morong21,Scherg21} would be natural.


\section*{Acknowledgments}
This work was supported by the National Natural Science Foundation of China (Grant No.~12604091), the High Level Talents Research and Startup Foundation Projects for Doctors of Zhoukou Normal University (Grant No.~ZKNUC2023016), the Natural Science Foundation Project of Henan Province of China (Grant No.~242300420941, No.~252300423541, No.~262300422459), the Research Project of Hubei Education Department (Grant No.~B2023077), and the Doctoral Scientific Research Foundation of Hubei University of Automotive Technology (Grant No.~BK202432).

\begin{appendix}

\section{Parameters and numerical procedures}
\label{app:num}

The three inputs \(\Fo\), \(t\) and \(\dt\) are fixed self-consistently by the three requirements \(|A(0)|=53.9795\)~Hz (the exact value, against the first-order estimate \(53.9995\)~Hz), \(\Dc^{(1)}=-235.8978\)~Hz and \(\Dc=-263.1889\)~Hz---a nonlinear system solved by Newton's method from a coarse grid of starting points, giving Table~\ref{tab:param}. 
The number of digits matters: rounding the three targets to \(53.980\), \(-235.9\) and \(-263.2\)~Hz displaces the solution to \(\Fo=567.94\)~Hz, \(t=226.573\)~Hz and \(\dt=-47.4124\)~Hz, far more than the digits of Table~\ref{tab:param} can absorb. 
Uncertainties in the last digit of \(\dt\) shift \(\Dc\) by less than \(0.1\)~Hz. All diagonalizations use dense Hermitian eigensolvers in double precision; the noise floor on the pairing residual is \(10^{-11}\)--\(10^{-10}\)~Hz. 
The amplitude is extracted from a finite chain without fitting, from the pairing plateau of Eq.~(\ref{eq:plateau}) for odd \(N\)---with \(n\) and \(O-n\) both at least four sites from either edge, where \(|\varepsilon_n|\) is below the \(10^{-3}\)~Hz level per the bound of Sec.~\ref{sec:finite}---and from the corrected alternating spacings of Eq.~(\ref{eq:Afromspacing}) for even \(N\); the two agree to \(10^{-11}\)~Hz. 

The generating scripts for every numerical result are provided as machine-readable supplementary material: \texttt{supp\_closed\_form.py} (Floquet propagator behind Eq.~(\ref{eq:Afloq})), \texttt{supp\_ws\_series.py} (WS recursion behind Eq.~(\ref{eq:series})), \texttt{supp\_ed\_verification.py} (finite-chain diagonalization), \texttt{supp\_magnus\_crosscheck.py} (independent Magnus integration), \texttt{supp\_resummation.py} (truncated series and in-window Pad\'e resummation), \texttt{supp\_pade.py} (converged-coefficient Pad\'e resummation of the remote zeros), \texttt{supp\_perturbations.py} (perturbation robustness checks of Sec.~\ref{sec:scope}), \texttt{supp\_sc\_mapping.py} (Schwarz--Christoffel mapping), \texttt{supp\_protocols.py} (readout protocols and error budget), and \texttt{supp\_figures.py} (figure generation). 
A README file (\texttt{supp\_README.md}) lists the integration steps, random seeds, chain lengths \(N\), tolerances, and the mapping from each figure of the paper to the script that generates it. We have explicitly checked that the results quoted in the text are stable under increasing \(N\): the zeros of \(A\) show no drift between \(N=40\) and \(1601\), and the pairing residual stays
in the \(10^{-11}\)--\(10^{-9}\)~Hz floor for \(N\ge40\).

The self-consistent fixing used above should not be mistaken for an assumption built into the series. 
As an independent check, we repeated the convergence test at two parameter sets chosen without reference to the exact zero: \((\Fo,t,\dt)=(1000,150,-30)\)~Hz and \((413,100,-15)\)~Hz. 
The first set keeps the working-point ratio \(\dt/t=-0.2\) while shifting \(\Fo\) and \(t\) to a different scale; the second changes the ratio to \(\dt/t=-0.15\). 
The dimensionless field \(\beta=2t/\Fo\) then takes the values \(0.300\) and \(0.484\), well below the working-point value \(0.798\). 
At each set the exact \(\Dc\) is determined a posteriori from the Floquet construction of Sec.~\ref{sec:floquet}, and the parameters are not adjusted to it. Table~\ref{tab:ind} gives the resulting zero locations. 
The error pattern mirrors that of the working point (Fig.~\ref{fig:fig2}(a))---first order carries tenths of a hertz, the third order reaches \(10^{-3}\)~Hz, and the fifth and seventh orders reach \(10^{-4}\)~Hz or better---and, because both independent points lie deeper in the strong-field regime, the series there converges \emph{faster} than at the working point. 
The sub-\(0.01\)~Hz precision of the seventh-order truncation is thus a property of the series itself, not of the self-consistent parameter fixing.

\begin{table}[tb]
\caption{\label{tab:ind} Location of \(\Dc\) by the analytic series at
increasing truncation order for two parameter sets chosen independently
of the exact zero (Sec.~\ref{sec:wsbasis}). Error is \(\Dc^{(n)}-\Dc\).}
\begin{ruledtabular}
\footnotesize
\begin{tabular}{lrlrl}
truncation & \multicolumn{1}{c}{\(\Dc\) (Hz)} & error (Hz) & \multicolumn{1}{c}{\(\Dc\) (Hz)} & error (Hz)\\
\hline
& \multicolumn{2}{c}{Set 1} & \multicolumn{2}{c}{Set 2}\\
\hline
first order & \(-37.7236\) & \(+0.0834\) & \(-33.0977\) & \(+0.1536\)\\
\(+\)third order  & \(-37.8071\) & \(-0.0001\) & \(-33.2518\) & \(-0.0005\)\\
\(+\)fifth order  & \(-37.8070\) & \(+0.0000\) & \(-33.2513\) & \(+0.0000\)\\
\(+\)seventh order& \(-37.8070\) & \(+0.0000\) & \(-33.2513\) & \(+0.0000\)\\
exact \(\Dc\)     & \(-37.8070\) & --- & \(-33.2513\) & ---
\end{tabular}
\end{ruledtabular}
\end{table}

\section{Wannier--Stark representation}
\label{app:ws}

The WS states of the untilted uniform chain are \(\ket{\psi_m}=\sum_n J_{n-m}(\beta)\ket{n}\) with \(\beta=2t/\Fo\). 
They are orthonormal and complete by the Bessel identities~\cite{Stey73,Fukuyama73}. The tilt is diagonal in them, \(\Braket{\psi_m}{\Fo\hat n}{\psi_{m'}}=m\Fo\delta_{mm'}\): the recursion \(2\nu J_\nu(x)=x[J_{\nu-1}(x)+J_{\nu+1}(x)]\) gives \(\bra{\psi_m}\hat n\ket{\psi_{m'}}=m\delta_{mm'}+ \frac{\beta}{2}(\delta_{m,m'+1}+\delta_{m,m'-1})\), and the two off-diagonal terms are precisely cancelled by the uniform hopping, whose matrix elements in the same basis are \(-t(\delta_{m,m'+1}+\delta_{m,m'-1})\): the off-diagonal part of \(\Fo\hat n\) is \(t(\delta_{m,m'+1}+\delta_{m,m'-1})\), so the sum vanishes.

The two modulations are not diagonal. 
The relevant identity is the Neumann addition theorem \(\sum_k J_k(x)J_{p-k}(y)=J_p(x+y)\), which with \(p=0\), \(y=x\) and \(J_{-k}=(-1)^kJ_k\) yields the staggered sum
\begin{equation}
\sum_n(-1)^nJ_{n-m}(x)J_{n-m'}(x)=(-1)^{m}J_{m-m'}(2x).
\label{eq:stag}
\end{equation}
With \(x=\beta\) the onsite modulation contributes \(\frac{\Delta}{2}(-1)^{m+s}J_s(2\beta)\), which is already of the form \(\frac12(-1)^{m+s}G_s\) with \(G_s^{\rm stag}=\Delta J_s(2\beta)\). 
The bond modulation couples \(\ket{n}\) to \(\ket{n\pm1}\); the two hoppings of a cell carry opposite staggering, and applying Eq.~(\ref{eq:stag}) to the shifted indices, then differentiating with respect to \(x\) and using \(2J_\nu'(x)=J_{\nu-1}(x)-J_{\nu+1}(x)\), gives the bond weight \(2\dt[J_{s-1}(2\beta)-J_{s+1}(2\beta)]=4\dt J_s'(2\beta)\), where the prime denotes the derivative with respect to the argument. 
Together these two terms give Eq.~(\ref{eq:wsme}) with \(G_s=\Delta J_s(2\beta)+4\dt J_s'(2\beta)\): the phase \((-1)^{m+s}\) in Eq.~(\ref{eq:wsme}) is the full phase of the matrix element, and no additional phase is absorbed into \(G_s\). 
The third-order amplitude built from these elements is used in the maximum-uniformity analysis of Sec.~\ref{sec:floquet}. 
We have checked Eq.~(\ref{eq:wsme}) against brute-force construction of the WS states for eight parameter sets spanning \(\Fo=300\)--\(1200\)~Hz and \(\beta=0.378\)--\(1.511\); the largest deviation is at the floating-point floor, below \(10^{-12}\)~Hz, i.e., below \(10^{-14}\) in relative terms measured against the largest matrix element of \(V\).

\section{All-orders strong-field series}
\label{app:series}

Write \(H=H_0+V\) with \(H_0=\Fo\sum m\ket{\psi_m}\bra{\psi_m}\) and \(V\) the off-diagonal (and \(s=0\) staggered) part of Eq.~(\ref{eq:wsme}).
Since \(G_s\) is explicit, the Rayleigh--Schr\"odinger recursion can be carried out symbolically. Writing the amplitude as in Eq.~(\ref{eq:series}), the first few orders are
\bea
A^{(1)}&=&-\frac{\Delta}{2}J_0(2\beta)+2\dt J_1(2\beta) ,\\
A^{(3)}&=&-\frac{1}{8\Fo^2}\sum_{s,t\ne0}\frac{(-1)^sG_sG_{t-s}G_t}{st}
+\frac{G_0}{8\Fo^2}\sum_{s\ne0}\frac{G_s^2}{s^2} ,
\label{eq:A3}
\eea
and \(A^{(5)}\), \(A^{(7)}\) follow from the same recursion. 
Let \(V_{i,i+s}\) denote the matrix elements of Eq.~(\ref{eq:wsme}), the diagonal \(V_{ii}=\tfrac12(-1)^iG_0\) included in the sums, and let \(d_i=i\Fo\); the WS energy corrections of two reference states \(i=0\) and \(i=1\) obey
\bea
E_i^{(m)}&=&\sum_{k\neq i}V_{ik}\,c_k^{(m-1)}\,,\label{eq:rsrec}\\
c_k^{(m)}&=&\frac{1}{d_i-d_k}\left[\sum_{n\neq i}V_{kn}\,c_n^{(m-1)}
-\sum_{p=1}^{m-1}E_i^{(p)}\,c_k^{(m-p)}\right]\,,\nonumber
\eea
(the second relation for \(k\neq i\)), starting from \(c_k^{(1)}=V_{ki}/(d_i-d_k)\); the amplitude coefficients are the half-difference \(A^{(m)}=-(E_0^{(m)}-E_1^{(m)})/2\) of the two reference levels (the relation is fixed at first order by \(A^{(1)}=-G_0/2\) and preserved by the recursion because the two levels are mapped onto each other by the joint inversion \((\Delta,\dt)\to(-\Delta,-\dt)\)). 
The coefficients are listed in Table~\ref{tab:coef}. (i) \(A^{(1)}\) reproduces the strong-field result of Ref.~\cite{MBK15}. 
Their Eqs.~(26) and~(27) express the first-order offset as a combination of \(J_0(z)\) and \(J_1(z)\) at \(z=2(J_1+J_2)/\Fo\), that is of \(J_0(2\beta)\) and \(J_1(2\beta)\), with coefficients \(\Delta/2\) and \(2\dt\); the two overall signs are set by which sublattice carries \(+\delta\) and which of the two bonds is called \(J_1\), both physically arbitrary, and with the natural choice their expression is Eq.~(\ref{eq:A1}) term by term. 
Evaluating their Eq.~(26) as a period average---an independent route, quadrature instead of Bessel evaluation---reproduces our closed form to \(1.2\times10^{-14}\)~Hz, one double-precision unit on the largest term (\(\sim60\)~Hz). 
(ii) On the line \(\dt=0\) our \(A^{(3)}\) has the structure of the third-order term of Ref.~\cite{MBK15}---cubic in \(\Delta\), proportional to \(\Fo^{-2}\), with the same Bessel argument---but that coefficient is given there only as a quadrature and has not been compared term by term. (iii) The fifth- and seventh-order coefficients have no counterpart in Ref.~\cite{MBK15}; they were re-derived by an independent implementation of the recursion, which reproduces every printed coefficient of Table~\ref{tab:coef} and the zeros of each truncation to \(10^{-6}\)~Hz.

\begin{table}[tb]
\caption{\label{tab:coef} Expansion coefficients of Eq.~(\ref{eq:series})
at \(\Fo=568\)~Hz. Units: \(c_{mk}\) in Hz\(^{1-m}\). Bracketed values
from a global fit to the numerical amplitude.}
\begin{ruledtabular}
\begin{tabular}{lr@{\hspace{3mm}}lr@{\hspace{3mm}}l}
\(c_{10}\) & \(-0.2289107\) & \(c_{11}\) & \(+1.138942\)\\
\(c_{30}\) & \(+3.706205\times10^{-7}\) & \(c_{31}\) & \(+8.742\times10^{-10}\)\\
        & \([\,3.706\times10^{-7}\,]\) & & \([\,8.7\times10^{-10}\,]\)\\
\(c_{32}\) & \(-3.296052\times10^{-7}\) & \(c_{33}\) & \(-1.832782\times10^{-7}\)\\
        & \([\,-3.296\times10^{-7}\,]\) & & \([\,-1.833\times10^{-7}\,]\)\\
\(c_{50}\) & \(+1.1409735\times10^{-13}\) & \(c_{51}\) & \(-2.8221428\times10^{-12}\)\\
\(c_{52}\) & \(+4.8447786\times10^{-12}\) & \(c_{53}\) & \(+7.3445994\times10^{-13}\)\\
\(c_{54}\) & \(-4.6961975\times10^{-12}\) & \(c_{55}\) & \(-1.8617435\times10^{-12}\)\\
\(c_{70}\) & \(-3.928719\times10^{-19}\) & \(c_{71}\) & \(-2.177231\times10^{-20}\)\\
\(c_{72}\) & \(+2.454304\times10^{-17}\) & \(c_{73}\) & \(-7.908043\times10^{-17}\)\\
\(c_{74}\) & \(+6.849169\times10^{-17}\) & \(c_{75}\) & \(+4.224148\times10^{-17}\)\\
\(c_{76}\) & \(-6.290231\times10^{-17}\) & \(c_{77}\) & \(-2.801636\times10^{-17}\)
\end{tabular}
\end{ruledtabular}
\end{table}

The vanishing of the even orders is a selection rule. 
Under \((\Delta,\dt)\to(-\Delta,-\dt)\) we have \(G_s\to-G_s\), so every term in the recursion containing an odd total number of \(G\)'s changes sign while the energy denominators, which depend on \(\Fo\) alone, do not. 
Since the staggered amplitude is the half-difference of two levels related by that transformation, it is odd, and the expansion contains only terms with \(m\) odd.\footnote{``\(M\)-th order'' denotes the truncation of Eq.~\eqref{eq:series} that keeps every term of odd total degree \(m\le M\); by the selection rule above these are the sectors \(m=1,3,\dots,M\), so fifteenth order carries eight sectors and not fifteen terms. 
This is not the one-dimensional Taylor expansion of \(A\) in \(\Delta\) at fixed \(\dt\): the latter has non-vanishing even-power coefficients---they collect the odd-total-degree terms carrying even powers of \(\Delta\), for instance \((m,k)=(3,1)\)---so truncating it at \(\Delta^{p}\) mixes total degrees \(p,p+2,p+4,\dots\) and corresponds to no finite \(M\).} 
Numerically the recursion returns exact zeros for the even orders, at the floating-point floor.

\begin{table}[tb]
\caption{\label{tab:domain} RMS and maximum error (Hz) of the truncated
series [Eq.~(\ref{eq:series})] against the Floquet reference
[Eq.~(\ref{eq:Afloq})] on three regions of the \((\Delta,\dt)\) plane.
Upper row: RMS; lower row: maximum. Regions: P, \(|\Delta|\le275\)~Hz;
D, \(|\dt|\in[70,110]\)~Hz; W, \(320\le|\Delta|\le520\)~Hz. Sampling:
\(11\times11\) grid.}
\begin{ruledtabular}
\begin{tabular}{llcccc}
region & definition & 1st & 3rd & 5th & 7th\\
\hline
P & \(|\Delta|\le275\), \(|\dt|\le50\) & \(3.85\) & \(0.318\) & \(0.0246\) & \(0.0037\)\\
  &                                & \(8.82\) & \(1.32\)  & \(0.100\)  & \(0.0166\)\\
D & \(|\dt|\in[70,110]\), \(|\Delta|\le300\) & \(5.49\) & \(1.58\) & \(0.477\) & \(0.125\)\\
  &                                & \(15.5\) & \(6.49\)  & \(2.38\)   & \(0.708\)\\
W & \(|\Delta|\in[320,520]\), \(|\dt|\le50\) & \(33.2\) & \(3.87\) & \(1.66\) & \(0.842\)\\
  &                                & \(63.4\) & \(11.7\)  & \(4.80\)   & \(3.75\)
\end{tabular}
\end{ruledtabular}
\end{table}

\section{Floquet propagator construction and verification}
\label{app:floquet}

Group sites into cells \(m=(2m,2m+1)\). After removing the cell tilt \(2m\Fo\), the cell Bloch Hamiltonian is
\bea
h(k)&=&\begin{pmatrix}
\Delta/2 & w(k)\\
w(k)^* & \Fo-\Delta/2
\end{pmatrix},\nonumber\\
w(k)&=&-(t+\dt)-(t-\dt)e^{-ik} .
\eea
In the acceleration frame the cell momentum obeys \(\dot k=-4\pi\Fo\) (the force on a cell is \(2\Fo\) because the cell is two sites long), so \(k(t)=k_0-4\pi\Fo t\) and the Bloch period of the cell is \(T=1/(2\Fo)\).
The propagator is \(U=\mathcal T\exp[-2\pi i\int_0^Th\,dt]\), evaluated with a fourth-order Magnus integrator using two-point Gauss--Legendre quadrature~\cite{Blanes09}. The convergence is fourth order---measured against an independent adaptive integrator of the same \(2\times2\) system, the error falls by a factor \(\simeq16\) per doubling of the step number---and reaches \(3\times10^{-12}\)~Hz already at \(4000\) steps.
Since \(\mathrm{tr}\,h=\Fo\) is constant, \(\det U=e^{-2\pi i\Fo T}=e^{\pi i}=-1\), whence \(\mathrm{Tr}\,U=2i\cos\theta\) and Eq.~(\ref{eq:Afloq}). 
The sign of \(\Delta\) entering \(h\) reflects the arbitrary choice of which sublattice is labelled \(A\) versus \(B\) in the folding; exchanging the labels sends \(\Delta\to-\Delta\) while leaving the spectrum invariant, so the sign has no physical content.

The naive prescription---driving the Rice--Mele Hamiltonian alone with \(k(t)=k_0+2\pi\Fo t\) over \(T=1/\Fo\)---fails already at \(\dt=\Delta=0\), where it returns \(\pm192.9\)~Hz instead of \(0\). 
There are two reasons: the true period is two sites, so the driving frequency is wrong by a factor of two; and folding the tilt into the cell basis generates a sublattice gauge offset (the \(-\Fo\) diagonal entry) together with the spinor boundary stitching, neither of which the bare Rice--Mele Hamiltonian contains.

We have verified the construction on five discriminating cases---the trivial limit, \(\Delta=0\), the line \(\dt=0\), the point \(\Dc\), and the second zero---finding agreement with diagonalization of the \(N=401\) chain to \(2\times10^{-11}\)~Hz---the coarser figure reflects the finite-chain benchmark, not the integrator---when the propagator is evaluated at \(1.2\times10^{5}\) steps per period, and confirming that the three zeros return \(A=0\), that \(|A|\le\Fo/2\) holds throughout, and that \(\mathrm{Tr}\,U=\pm2i\) at the four exceptional points.

\section{Singularities and resummation}
\label{app:sing}

In Eq.~(\ref{eq:Afloq}) the principal branch of the arcsine is understood on the real axis; the analytic structure in the complex plane is fixed by \(\mathrm{Tr}\,U\) itself. \(A(\Delta)\) is analytic on the real axis; its singularities are the exceptional points of \(U\), where the two
eigenvalues coalesce, i.e., \(\mathrm{Tr}\,U=\pm2i\). 
They are located by
minimizing \(|(\mathrm{Tr}\,U)^2+4|^2\) in the complex plane (multi-start
Nelder--Mead): a \(51\times51\) grid of seeds over
\(|\mathrm{Re}\,\Delta|,|\mathrm{Im}\,\Delta|\le1000\)~Hz converges to
exactly the four exceptional points listed below, with residuals below
\(10^{-16}\); within this domain there is no closer singularity. Solving
gives two conjugate pairs, \(\Delta_{\rm EP}=428.075\pm383.614i\)~Hz and
\(-510.326\pm562.712i\)~Hz. The nearest pair sets the radius of convergence
\(R=|\Delta_{\rm EP}|=574.811\)~Hz, consistent with the practical breakdown
of the truncated series in the range \(|\Delta|\simeq320\)--\(520\)~Hz:
there the series still converges, but too slowly for the tabulated orders
to be useful.

Pad\'e approximation of the Taylor coefficients~\cite{Baker96} extends the
reach: an \([18/18]\) approximant built from the recursion coefficients
remains accurate on the real axis out to \(|\Delta|\simeq500\)~Hz at the
\(10^{-7}\)~Hz level (\(4\times10^{-5}\)~Hz at \(|\Delta|\simeq600\)~Hz),
against \(\simeq320\)~Hz for the seventh-order truncation.

On the positive side, resummation does reach the nearer remote zero without
the Floquet construction: a Pad\'e approximant built from the Taylor
coefficients alone---equivalently from the recursion of
Appendix~\ref{app:series}---reproduces \(\Dcm\) to the \(0.1\)-Hz level
(\([10/10]\): \(0.12\)~Hz; \([19/6]\): \(10^{-3}\)~Hz) while \(\Dcp\) is
reproduced only erratically, from a few Hz to several hundred Hz and with
no convergence in the order. What fails is systematic accuracy, not
accessibility. On the negative side, the conformal map is not the operative
tool: that the inverse images of the zeros lie in the unit disk is true of
every point in the image, and the relevant series is that of \(A\circ f\)
in the disk variable, not the composition \(\sum_ka_kf(w)^k\), which by
construction converges only where \(|f(w)|<R\). A genuine resummation of
\(A\)---Borel--Pad\'e being one option~\cite{Berry90}---is what the remote
zeros call for. We have not pursued this, because the Floquet construction
(\ref{eq:Afloq}) already delivers the same information to \(10^{-4}\)~Hz.

\section{Odd chains: defect norm, bound, and plateau}
\label{app:odd}

\emph{Defect norm.} From Eq.~(\ref{eq:specinv}),
\(D=2(H_{\dt}+H_{\rm stag})\). In the site basis \(D\) is a signed
tridiagonal matrix plus a staggered diagonal. Write \(S\) for the signed
adjacency, \(S_{n,n+1}=S_{n+1,n}=(-1)^{n+1}\), so that
\(D=\Delta\,\mathrm{diag}[(-1)^n]+2\dt\,S\). The gauge
\(g_n=(-1)^{\lfloor n/2\rfloor}\) carries \(S\) into the unsigned
adjacency, \(gSg=-\mathcal{A}_{\rm adj}\); being a unitary (indeed
diagonal, \(g^2=\one\)) transformation it preserves the norm,
\(\|S\|=\|\mathcal{A}_{\rm adj}\|\), which is the only property of \(S\)
needed below. (The reflection, by contrast, satisfies
\(R\mathcal{A}_{\rm adj}R=+\mathcal{A}_{\rm adj}\) and is not what produces
the sign.) The two pieces of \(D\) anticommute---the anticommutator has
matrix elements \([(-1)^n+(-1)^{n+1}]\mathcal{A}_{n,n+1}=0\)---and
\be
\|D\|^2=\Delta^2+16\dt^2\|\mathcal{A}_{\rm adj}\|^2
=\Delta^2+16\dt^2\cos^2\frac{\pi}{N+1} ,
\ee
using \(\|\mathcal{A}_{\rm adj}\|=2\cos[\pi/(N+1)]\) for a path of \(N\)
sites. At \(\Delta=\Dc\) this is \(324.399\)~Hz in the large-\(N\) limit;
diagonalization gives \(324.197\) (\(N=51\)), \(324.347\) (\(101\)),
\(324.396\) (\(401\)) and \(324.399\)~Hz (\(1001\)), matching the Floquet
value to all digits.

\emph{Bound.} Since \(\Sigma\) is unitary, \(H\) and
\(\Sigma H\Sigma^\dg=\Fo O\one-H+D\) have the same spectrum. Ordering
eigenvalues ascending, \(\lambda_n(\Fo O\one-H)=\Fo O-E_{N-1-n}\) and
\(\lambda_n(\Fo O\one-H+D)=E_n\), so Weyl's inequality
\(|\lambda_n(M_2)-\lambda_n(M_1)|\le\|D\|\) gives
Eq.~(\ref{eq:oddbound}).

\emph{Plateau.} At first order the two sites \(n\) and \(O-n\) of an odd
chain lie on the same sublattice (with \(\eta_n\) as in
Eq.~\eqref{eq:eta}), so their staggering offsets add and the pairing
residual is \(2(-1)^{n+1}A^{(1)}\). Beyond first order the plateau value
could in principle receive corrections that are odd under the joint
inversion without being proportional to \(A^{(1)}\)---for example, \(A\)
times an even function of \(\Delta\) and \(\dt\). We therefore do not claim
an all-orders proof: the plateau stands as a proposition supported by the
first-order argument and by numerical verification. Numerically the plateau
matches \(2|A|\) from Eq.~(\ref{eq:Afloq}) to \(1.3\times10^{-7}\)~Hz with
a spread below \(10^{-9}\)~Hz across \(N=21,\dots,201\)
(Table~\ref{tab:plateau}), and it is independent of \(N\).

\begin{table}[tb]
\caption{\label{tab:plateau} Odd-\(N\) pairing plateau versus \(2|A|\)
from the Floquet construction for several offsets. Agreement limited by
the propagator mesh (\(10^{-7}\)~Hz); diagonalization noise floor
\(10^{-11}\)--\(10^{-10}\)~Hz (App.~\ref{app:num}).}
\begin{ruledtabular}
\begin{tabular}{lccc}
\(\Delta\) (Hz) & plateau (Hz) & \(2|A|\) (Hz) & difference (Hz)\\
\hline
\(-500\) & \(38.21\) & \(38.21\) & \(<1.5\times10^{-7}\)\\
\(-200\) & \(21.8518\) & \(21.8518\) & \(<1.5\times10^{-7}\)\\
\(0\)    & \(107.96\) & \(107.96\) & \(<1.5\times10^{-7}\)
\end{tabular}
\end{ruledtabular}
\end{table}

\emph{Edge correction.} The residual after subtracting the plateau is
localized at the ends. Table~\ref{tab:odd} compares \(|\varepsilon_d|\)
with the bound \(4tJ_d(2t/\Fo)^2\); the bound is satisfied with margin
across \(\Delta\in[-500,500]\)~Hz. This is a bound and not a scaling law:
the measured local slope oscillates between odd and even \(d\), because the
two sublattices contribute with opposite signs. A single-exponential fit
over \(d=1,\dots,5\) at \(\Delta=\Dc\) (\(N=401\)) gives an effective decay
length \(0.22\) sites, against the WS localization length
\(\beta/\sqrt2=0.56\) sites (rms) of Sec.~\ref{sec:model}: the edge
correction decays several times faster than the states it corrects. Both
numbers are effective parameters---the fitted length depends on the fit
range, and the true decay is faster than exponential (Bessel factorial
tail).

\begin{table}[tb]
\caption{\label{tab:odd} Ratio \(|\varepsilon_d|/[4tJ_d(2t/\Fo)^2]\) for
\(d=0,\dots,5\), maximized over \(\Delta\in[-500,500]\)~Hz on an \(N=401\)
chain. The bound is never violated.}
\begin{ruledtabular}
\begin{tabular}{lcccccc}
\(d\) & \(0\) & \(1\) & \(2\) & \(3\) & \(4\) & \(5\)\\
\hline
ratio & \(0.369\) & \(0.255\) & \(0.119\) & \(0.163\) & \(0.088\) & \(0.120\)
\end{tabular}
\end{ruledtabular}
\end{table}

\section{Readout protocols}
\label{app:protocols}

\emph{Protocol A.} Equation~(\ref{eq:Afloq}) is fixed by its calibrated
inputs alone, so \(A(\Delta)\) is reconstructed over the whole scanned range
without additional fitted parameters. The assertion to be tested is the
simultaneous reproduction of the three zero positions,
\(\Dcm=-710.0978\), \(\Dc=-263.1889\) and \(\Dcp=+920.8092\)~Hz, and of the
window maximum \(|A|=117.218\)~Hz against the bound \(\Fo/2=284\)~Hz.

\emph{Protocol B.} A wave packet is evolved for \(T_2\) and the sublattice
imbalance is band-passed about \(\Fo\) and Fourier transformed. The sideband
amplitude \(M\) is linear in \(A\); the proportionality constant is set by
the packet width and by the normalization of the imbalance signal, and is
held fixed throughout, so that only the linearity---not the
constant---enters the protocol. We simulated the full protocol including
the finite evolution window and find that the readout is linear, but that
the curvature of \(A(\Delta)\) over the accessible window
\(|\Delta-\Dc|\lesssim120\)~Hz biases a linear extrapolation by
\(\approx-9\)~Hz (\(-9.48\)~Hz for a \(20\)~Hz grid, varying between
\(-9.5\) and \(-8.7\)~Hz for \(20\)--\(10\)~Hz steps; the sign assumes a
two-sided scan). A shape-constrained fit using Eq.~(\ref{eq:Afloq}) removes
the bias; a straight-line extrapolation does not. The linewidth criterion
\(4|A|\sim1/T_2\) translates, through the measured slope \(0.3148\) of
\(2|A|\) versus \(|\Delta-\Dc|\) [Fig.~\ref{fig:fig6}(b)], into onset
thresholds \(|\Delta-\Dc|\simeq15.9\), \(5.3\) and \(1.6\)~Hz for
\(T_2=0.1\), \(0.3\) and \(1.0\)~s.

\emph{Tilt-and-setpoint budget.} This is the tilt-and-setpoint part of
Eq.~(\ref{eq:propagation}), evaluated in Sec.~\ref{sec:error}:
\(\delta\Dc=[(1.5255\,\delta\Fo)^2+\delta\Delta^2]^{1/2}\). Here
\(\partial\Dc/\partial\Delta=1\) because an error \(\delta\Delta\) of the
detuning setpoint shifts the inferred zero one-to-one---\(\Dc\) is the zero
of \(A(\Delta)\), so the derivative is with respect to the setpoint, not to
a root variable---while \(\partial\Dc/\partial\Fo=1.5255\) dominates. The
budget carries \(\delta\Fo\) and \(\delta\Delta\) only; \(\delta\,t\) and
\(\delta\dt\) enter through the full propagation of Sec.~\ref{sec:error}.

\section{Proof of the hidden symmetry}
\label{app:sym}

With \(\Gamma\ket{n}=(-1)^n\ket{n}\) and \(P\ket{n}=\ket{O-n}\) one has
\(P\Gamma=(-1)^O\Gamma P\), so \(\Sigma^2=(-1)^O\one\) while
\(\Sigma^\dg\Sigma=\one\) always; \(\Sigma\) is Hermitian for odd \(N\) and
anti-Hermitian for even \(N\). To verify Eq.~(\ref{eq:specinv}), note that
\(\Sigma\) transports operators by reflection,
\(\Sigma\ket{n}\bra{m}\Sigma\dg=(-1)^{n+m}\ket{O-n}\bra{O-m}\) (the two
phases come from \(\Sigma\ket{n}=(-1)^n\ket{O-n}\) and its adjoint). A bond
term \(\ket{n}\bra{n+1}\) is therefore carried into
\(-\ket{O-n}\bra{O-n-1}\); relabelling \(n\to O-1-n\) identifies this as
minus the bond term of the reflected cell, and the factor \(-1\) is
independent of \(n\) and of \(O\). An onsite term \(\ket{n}\bra{n}\) is
carried into \(+\ket{O-n}\bra{O-n}\). Consequently the uniform hopping and
the tilt transform homogeneously---giving the \(\Fo O\one-H\) part of
Eq.~(\ref{eq:specinv})---while the two staggered pieces, whose amplitudes
alternate in sign, are carried into \((-1)^O\) times themselves rather than
minus themselves: they acquire the extra factor \([1+(-1)^O]\), which
vanishes for even \(N\) and equals \(2\) for odd \(N\). This is
Eq.~(\ref{eq:specinv}).

For even \(N\), \(\Sigma H\Sigma^\dg=\Fo O\one-H\) implies that the spectrum
is symmetric about \(\Fo O/2\) and that Eq.~(\ref{eq:pairing}) holds after
ordering: the reflection \(n\mapsto O-n\) has no fixed point since \(O\) is
odd, so the \(N\) eigenvalues are exhausted by disjoint pairs. Combining
with Eq.~(\ref{eq:ladder}), \(E_n+E_{O-n}=\Fo O\) gives \(2C=0\) upon
summing the constant offsets, so \(C\equiv0\) identically in \(\Delta\) and
\(\dt\)---not merely at some special point. The joint parity
\(A(-\Delta,-\dt)=-A(\Delta,\dt)\) is a separate statement and does not
follow from \(\Sigma\): it comes from the sublattice exchange of the bulk,
i.e., from translation by one site, which reverses both modulations,
\(n\mapsto n+1: (\Delta,\dt)\mapsto(-\Delta,-\dt)\), and shifts the tilt by
the constant \(\Fo\), absorbed by relabelling the ladder index in
Eq.~(\ref{eq:ladder}); the staggered amplitude therefore changes sign while
every level spacing is unchanged. (Conjugation by \(\Gamma\) alone is not
the right operation: it flips the sign of \emph{all} hoppings, uniform and
staggered alike, and leaves both diagonal terms untouched.)

\end{appendix}

\bibliography{refs}

\end{document}